%% file: manuscript-jt.tex
\documentclass[conference]{IEEEtran}
\IEEEoverridecommandlockouts
\usepackage{cite}
\usepackage{amsmath,amssymb,amsfonts}
\usepackage{algorithmic}
\usepackage{graphicx}
\usepackage{textcomp}
\usepackage{booktabs}
\usepackage{threeparttable}
\usepackage{tabularx}
\usepackage{xcolor}
\usepackage{multirow}

\usepackage{tikz}
\usepackage{textcomp}
\usepackage{hyperref}
\usepackage{lipsum}

\newcommand\copyrighttext{%
  \footnotesize \textcopyright 2024 IEEE. Personal use of this material is permitted.
  Permission from IEEE must be obtained for all other uses, in any current or future
  media, including reprinting/republishing this material for advertising or promotional
  purposes, creating new collective works, for resale or redistribution to servers or
  lists, or reuse of any copyrighted component of this work in other works.
  DOI: 10.1109/CogMI62246.2024.00030}
\newcommand\copyrightnotice{%
\begin{tikzpicture}[remember picture,overlay]
\node[anchor=south,yshift=10pt] at (current page.south) {\fbox{\parbox{\dimexpr\textwidth-\fboxsep-\fboxrule\relax}{\copyrighttext}}};
\end{tikzpicture}%
}

\def\BibTeX{{\rm B\kern-.05em{\sc i\kern-.025em b}\kern-.08em
    T\kern-.1667em\lower.7ex\hbox{E}\kern-.125emX}}
\begin{document}

\title{Advancing Mental Health Pre-Screening: A New Custom GPT for Psychological Distress Assessment\\}

\author{
\IEEEauthorblockN{Jinwen Tang and Yi Shang}
\IEEEauthorblockA{\textit{Electrical Engineering and Computer Science Department} \\
\textit{University of Missouri}\\
Columbia, Missouri, USA \\
\{jt4cc, shangy\}@umsystem.edu}
}

\maketitle
\copyrightnotice

\begin{abstract}
This study introduces 'Psycho Analyst', a custom GPT model based on OpenAI's GPT-4, optimized for pre-screening mental health disorders. Enhanced with DSM-5, PHQ-8, detailed data descriptions, and extensive training data, the model adeptly decodes nuanced linguistic indicators of mental health disorders. It utilizes a dual-task framework that includes binary classification and a three-stage PHQ-8 score computation involving initial assessment, detailed breakdown, and independent assessment, showcasing refined analytic capabilities. Validation with the DAIC-WOZ dataset reveals F1 and Macro-F1 scores of 0.929 and 0.949, respectively, along with the lowest MAE and RMSE of 2.89 and 3.69 in PHQ-8 scoring. These results highlight the model's precision and transformative potential in enhancing public mental health support, improving accessibility, cost-effectiveness, and serving as a second opinion for professionals.
\end{abstract}

\begin{IEEEkeywords}
Mental Health, LLM Application, Generative AI, Psychological Distress Assessment
\end{IEEEkeywords}

\section{Introduction}
The development of Generative Pre-trained Transformers (GPTs) has marked a significant advancement in artificial intelligence \cite{openai2023gpt}. 
In the realm of public health, mental well-being has become increasingly recognized as a crucial aspect, yet it faces challenges in terms of recognition and accessibility. Societal barriers, including stigma and reluctance to discuss psychological distress, hinder individuals from seeking timely and effective clinical intervention \cite{gulliver2010perceived}\cite{pescosolido1999people}. These challenges are further compounded by economic constraints and limited availability of mental health services. In this context, the advancements of Large Language Models (LLMs), such as GPT-3.5 and GPT-4, offer new possibilities for discreet, accessible, and non-judgmental mental health support.

Our study explores the potential of GPT-4 and specialized GPT variants in mental health research, particularly focusing on their ability to facilitate empathetic and nuanced interactions that are conducive to preliminary mental health screenings and emotional support. 
We implemented 'Psycho Analyst', a custom GPT model specifically designed for mental health pre-screening through textual analysis. 
We conducted a comprehensive evaluation of this model using the DAIC-WOZ database \cite{devault2014simsensei}, measuring its efficacy in identifying current mental health issues, and  demonstrating the model's high precision. 
These results not only highlight the capability of Psycho Analyst GPT in detecting mental health concerns but also mark a crucial step in early intervention strategies.
Furthermore, the Psycho Analyst GPT's ability to analyze extensive data from clinical dialogues and studies positions it as an essential tool for discerning trends and patterns in mental health, thereby enriching research and understanding of various psychological conditions. The model's ability to be fine-tuned with specific therapeutic approaches and psychological theories paves the way for more individualized therapeutic interventions.

This paper contributes to the field in several areas:
\begin{enumerate}
\item \textbf{Innovative Model Development:} We developed 'Psycho Analyst,' a custom GPT model optimized for mental health pre-screening, utilizing OpenAI's ChatGPT-4 service. This model uniquely integrates the DSM-5 and PHQ-8, both of which are globally recognized for their clinical validity. Additionally, it incorporates detailed data descriptions and training data from the DAIC-WOZ database. This comprehensive integration significantly enhances the model’s capacity to accurately interpret nuanced language, enabling it to effectively identify various mental health conditions.
\item \textbf{Dual Task Framework:} 'Psycho Analyst' operates through a sophisticated dual-task framework designed to handle both classification of mental health status and computation of PHQ-8 scores. This dual approach allows for a more comprehensive evaluation of mental health conditions, offering both binary classification and detailed severity assessments.
\item \textbf{Empirical Evaluation and Validation:} The Psycho Analyst model has been rigorously validated using the DAIC-WOZ clinical transcript dataset. Our evaluations confirm the model's effectiveness in real-world clinical environments, reinforcing its practical applicability for mental health diagnostics. This empirical assessment not only underscores the robustness of Psycho Analyst but also highlights its superior performance in comparison to other large language models such as GPT-4o and Mixtral-8*7B.
\item \textbf{Innovative Application in Learning Scenarios:} The model's superior performance in zero-shot and few-shot learning scenarios highlights its capability to adapt to varied data with minimal training. This feature is particularly valuable in mental health settings where the model can quickly adjust to the nuances of different patient interactions without extensive retraining.
\item \textbf{New potential for Public Mental Health:} By successfully integrating DSM-5 and PHQ-8 criteria into the generative AI analysis process, the Psycho Analyst model opens new possibilities for early intervention and tailored mental health care. This integration enables more precise, personalized, and cost-effective assessments, significantly improving the potential for early detection and customized treatment plans in public health settings.
\end{enumerate}

In summary, this work underscores the advanced capabilities of custom GPT models in mental health diagnostics and heralds new pathways for accessible and personalized mental health care. 

In the rest of this paper, we will delve into the background and related work to provide context and foundation for our study. We will then introduce our data and methodology, detailing the development and configuration of the Psycho Analyst GPT model. This is followed by a presentation of our analyses and results, where we evaluate the model's performance. Finally, we will discuss the implications of our findings, addressing the limitations of our study, and exploring future research directions.

\section{Background and Related Work}
In the realm of mental health research, leveraging machine learning and advanced natural language processing methods has become increasingly prevalent. These technologies have been used to analyze a range of data types, including brain imaging, clinical notes, mobile sensor data, and surveys. Studies have aimed at identifying individuals at risk of suicide, classifying mental health conditions such as major depressive disorder (MDD) \cite{patel2015machine} and schizophrenia \cite{drysdale2017resting}, differentiating between disorders with similar symptoms \cite{erguzel2016artificial}, and predicting the severity of mental health issues \cite{kacem2018detecting}. These efforts signify a paradigm shift toward data-driven insights for enhancing mental health outcomes.

Despite these advancements, a significant challenge persists in effectively reaching and assisting the majority of individuals grappling with mental health concerns. Approximately two-thirds of such individuals are hesitant to seek professional help, with an even smaller fraction consulting mental health specialists \cite{andrews2001shortfall, regier1993facto}. This reluctance often stems from societal stigma and discomfort in openly discussing mental health issues.

In response to these challenges, recent research has increasingly focused on utilizing text data from social media platforms, including blogs, Twitter, Reddit, and Chinese Weibo \cite{zhang2022natural}. These platforms offer a rich, publicly accessible source of personal expressions, enabling researchers to detect and monitor mental health states more effectively \cite{graham2019artificial, shatte2019machine}. However, the majority of these studies do not directly involve clinical diagnoses, which limits the precision of their findings.

Traditionally, mental health research has relied heavily on Electronic Health Records (EHRs) and clinical notes \cite{le2021machine, tran2017predicting}. These sources, while rich in professional and structured content, often lack the natural expressions of patients, as they are typically summaries or rephrasings by healthcare professionals. Despite their value, access to these records is generally restricted, and they may not fully capture the personal linguistic nuances associated with various mental health conditions.

Recognizing the limitations of both social media data and traditional clinical records, the DAIC-WOZ dataset and its extensions \cite{gratch2014distress, devault2014simsensei, ringeval2019avec} provide a unique blend of narrative speech and clinical interview transcripts, offering insights into natural linguistic expressions coupled with clinical diagnoses. This approach allows for a more nuanced analysis of language related to anxiety, depression, and PTSD, bridging the gap between clinical accuracy and natural expression.

Villatoro et al. \cite{villatoro2021approximating} used a mental lexicon approach to differentiate between depressed and non-depressed individuals in clinical interviews, achieving macro F1 scores of 0.83 in the DAIC-WOZ and E-DAIC datasets.

Solieman et al. \cite{solieman2021detection} extended this research by diagnosing depression through audio recordings, using two models that processed data from the DAIC-WOZ database, yielding F1-scores of 0.8 and 0.76. Belser et al. \cite{belser2023comparison} compared three NLP models – BGRU, HAN, and Long-sequence Transformer – for screening depression in the DAIC-WOZ dataset, with both Transformer and HAN models achieving accuracies of 0.77 and F1-scores of 0.76.

Milintsevich et al. \cite{milintsevich2023towards} developed a multi-target hierarchical regression model for predicting individual depression symptoms from interviews in the DAIC-WOZ corpus, achieving a macro-F1 score of 73.9 and MAE of 3.78 for total depression score prediction.

Yadav et al. \cite{yadav2023novel} proposed a novel automated depression detection approach using linguistic content from patient interviews. This approach comprised a Bidirectional Gated Recurrent Unit (BGRU) network for linguistic information processing and a fully connected network to assess the depressed state. Validated using the Distress Analysis Interview Corpus-Wizard-of-Oz interviews dataset, the approach achieved an impressive F1 score of 0.92, outperforming previous models. This study highlighted the effectiveness of BGRU over Long Short Term Memory models in recognizing depression.

\section{Data Description and Preprocessing}
In this paper, we utilized the DAIC-WOZ dataset, a subset of the larger Distress Analysis Interview Corpus (DAIC) \cite{gratch2014distress}, extensively described in \cite{devault2014simsensei}. This dataset is distinguished by its integration of audio and video recordings alongside comprehensive questionnaire responses, obtained from Wizard-of-Oz style interviews. These interviews were facilitated by 'Ellie,' an animated virtual interviewer operated remotely by a human interviewer. The primary focus of the DAIC-WOZ database lies in its clinical interviews, which are specifically designed to facilitate the diagnosis of psychological distress conditions such as anxiety, depression, and PTSD.
\input{table1}

The dataset encompasses 187 interview entries that are divided into 105 records for training, 35 for validation, and 47 for testing. 
 Table~\ref{table:table1} shows the number of participants with PHQ-8 scores smaller than or equal or larger than 10, respectively, in different subsets of the DAIC-WOZ dataset. Patient Health Questionnaire-8 (PHQ-8) is a standard depression screening metric.
 
 \input{table2}
 
Table~\ref{tab:table2} shows sample excerpts of raw transcript from the interactions between the virtual interviewer--Ellie and the participants from the DAIC-WOZ dataset.

\begin{figure}[t]
  \centering
  \includegraphics[width=\columnwidth]{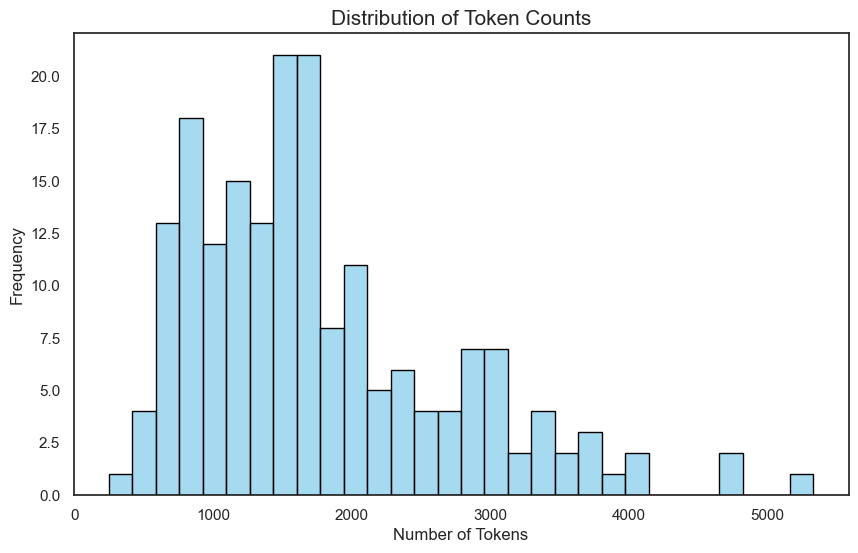} 
  \caption{Histogram of ChatGPT token counts of combined participant dialogues in the DAIC-WOZ dataset. }
  \label{fig:Token}
\end{figure}

To prepare the data for the development and evaluation of the Psycho Analyst model, we extracted textual responses from participants within the DAIC-WOZ dataset. These responses were concatenated to form a continuous text sequence for each participant. This method ensures that the context of each question and response is maintained, which is crucial for the model's understanding and analysis. To enhance the preservation of question context and minimize potential misinterpretations, we included markers ("./ ") to separate response within the concatenated sequences. This adjustment aims to provide clearer separations between distinct conversational turns, thereby aiding the model in its contextual comprehension. Table~\ref{tab:table3} shows samples of these concatenated dialogues of 10 participants. Moreover, the distribution of ChatGPT token counts of combined participant dialogues in the DAIC-WOZ dataset is depicted in Fig. \ref{fig:Token}.

\input{table3}

\section{Methods}
\subsection{Model Development}
We developed 'Psycho Analyst,' a custom GPT utilizing OpenAI's ChatGPT-4 service, tailored for mental health pre-screening. This model is designed to identify signs of mental health disorders such as anxiety, depression, stress, and bipolar within textual data. It integrates the Diagnostic and Statistical Manual of Mental Disorders, Fifth Edition (DSM-5), and the Patient Health Questionnaire-8 (PHQ-8) as foundational elements. These standards are globally recognized for their clinical validity and provide structured diagnostic criteria, thereby enhancing the model's capability to accurately interpret nuanced language indicative of mental health conditions.

\subsection{Dual Task Framework}
Psycho Analyst operated through structured prompts that guided its analytical processes. Its evaluation framework was based on two distinct tasks: (1) classification and (2) PHQ-8 score computation. Its performance was evaluated using various settings of background knowledge, as discussed later.

\subsubsection{Task 1: Classification}
To accurately reflect the complexities of mental health assessments, we used a nuanced 7-point scale for classifying mental health conditions. This scale ranged from 1, indicating 'not at all likely,' to 7, representing 'extremely likely.' This method allowed for a more detailed representation of the likelihood of a condition, accommodating the often nuanced and uncertain nature of mental health diagnostics. The final classification into binary outcomes was computed based on a threshold value determined through careful evaluation of the likelihood scores.

The single comprehensive prompt guided the model to:
\begin{itemize}
\item \textbf{Assess Mental Health Status:} Determine whether each patient, referenced by their ID, was likely to have mental health issues. This included a detailed analysis of the conditions present, using linguistic indicators found in the text data.
\item \textbf{Quantify Condition Severity:} Calculate the severity on a 7-point scale. This approach provided a more refined assessment compared to a simple binary outcome, better capturing the subtleties and uncertainties inherent in mental health evaluations.
\end{itemize}

This framework enhanced the model’s ability to conduct detailed and context-sensitive analyses, furthering the application of generative AI in mental health evaluations.

Table~\ref{tab:table4} presents examples of the model's outputs, illustrating how Psycho Analyst processed clinical interview transcripts to assess mental health conditions and assign likelihood scores. These examples emphasize the model's capability to move beyond mere yes/no classifications, thus accommodating the often ambiguous and complex nature of psychological symptoms for more clinically relevant and precise evaluations.
\input{table4}

\subsubsection{Task 2: PHQ-8 Score Computation}
\begin{figure*}[h!]
  \centering
  \includegraphics[width=1\linewidth]{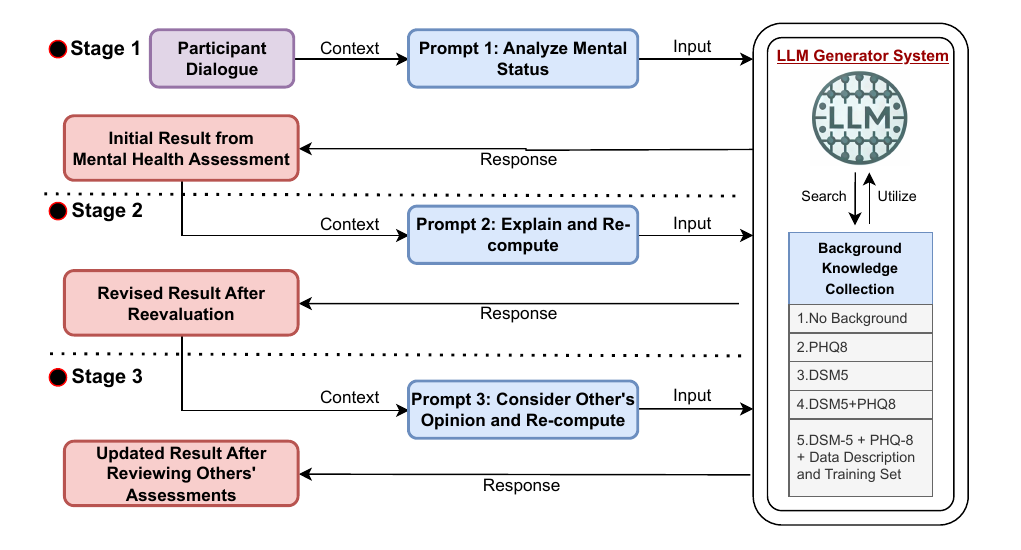}
  \caption{Three-Stage Mental Health Evaluation Framework: Comprehensive Analysis, Re-Evaluation, and Independent Validation of Others' Opinions.}
  \label{fig:jtllm_3}
\end{figure*}

The PHQ-8 assessment task, a comprehensive three-step process, evaluates how effectively our model, Psycho Analyst, analyzes mental health symptoms from individual transcripts. A detailed visual representation of this task is shown in Figure \ref{fig:jtllm_3}. The PHQ-8 tool measures mental health challenges on a scale from 0 to 24. Scores of 10 or higher might indicate significant mental health concerns, while scores of 20 or higher suggest more severe issues.

\paragraph{Stage 1: Initial Analysis}
In the first stage, Psycho Analyst is tasked with conducting a detailed analysis of a clinical interview transcript to determine the mental health status of an individual. This analysis is comprehensive, utilizing both the content of the transcript and the specific evaluation criteria outlined in the background materials provided. The evaluation focuses on two primary objectives:
\begin{itemize}
\item \textbf{Mental Health Assessment:} Analyze the transcript to determine if the individual has mental health issues, providing a reasoned explanation for the assessment. The analysis should cite relevant portions of the transcript and illustrate how these align with the criteria provided in the background knowledge, including the PHQ-8 items.
\item \textbf{PHQ-8 Score Estimation:} Estimate the PHQ-8 score for the individual based on the transcript's content and the evaluation guidelines from the background materials. The PHQ-8 score is crucial for assessing the severity of mental health symptoms.
\end{itemize}

\paragraph{Stage 2: Detailed Breakdown}
In the second stage, the model is asked to break down how it derives the PHQ-8 score for each symptom. This involves a step-by-step explanation of the scoring process, ensuring transparency and clarity in how the PHQ-8 score is computed based on the individual's responses in the transcript.

\paragraph{Stage 3: Independent Assessment}
The third stage evaluates the model's capacity for independent reasoning. Psycho Analyst is presented with a PHQ-8 score previously assigned to an individual by another evaluator, extracted from the original dataset. The tasks for the model in this stage are as follows:
\begin{itemize}
\item \textbf{Agree or Disagree:} Determine whether it agrees with the assigned PHQ-8 score.
\item \textbf{Explain Reasoning:} Provide a detailed explanation for its agreement or disagreement with the assigned score.
\item \textbf{Reanalyze and Discuss:} Reanalyze the PHQ-8 score, discussing the implications of the score and any potential discrepancies with the initial assessment.
\end{itemize}
Table~\ref{tab:table_3stage} shows an abbreviated simple output of this task.
\input{table_3stage}

\subsection{Evaluation}
Using the concatenated DAIC-WOZ dataset, we empirically evaluated Psycho Analyst's performance across different configurations of background knowledge, designed to simulate a variety of clinical knowledge settings. The configurations were as follows:
\begin{enumerate}
    \item A baseline scenario devoid of any background information, serving as a control.
    \item Incorporation of the PHQ-8, a widely-used depression screening tool consisting of eight questions that assess symptoms and severity of depression based on patient self-reporting. This configuration aims to evaluate the model's efficacy in applying standard depression screening metrics.
    \item Incorporation of the DSM-5, the definitive guide for psychiatric diagnoses. This integration includes comprehensive criteria and standardized language for mental disorders, intended to assess the model's alignment with widely recognized psychiatric criteria.
    \item A hybrid configuration with both PHQ-8 and DSM-5 to assess the synergistic effect of these diagnostic tools.
    \item An enhanced configuration that integrates DSM-5, PHQ-8, and a detailed data description from the dataset's curator along with an expanded training set, designed to optimize the model's contextual understanding and diagnostic precision.
\end{enumerate}
Additionally, Psycho Analyst's performance was compared with two baseline large language models: GPT-4o, a newly released model, and Mixtral-8*7B, a highly effective local model. Both models operated without any background knowledge, facilitating a direct assessment of Psycho Analyst's specialized enhancements. Mixtral-8*7B, running on an NVIDIA A100 GPU with 80GB of memory, provided a robust benchmark for computational efficiency and data processing capabilities. This comparison helped to contextualize the advancements of Psycho Analyst and its enhanced capability to interpret complex clinical data effectively. The effectiveness of each configuration was then assessed using the test set from the DAIC-WOZ dataset.

\section{ Results and Analysis}
This section evaluates the performance of the models for both tasks, beginning with an examination of execution times across different AI models.

Execution times varied significantly among the models tested. OpenAI's models, including GPT-4 and GPT-4o, processed Task 1 in approximately 10-20 seconds and completed the three-stage Task 2 in about 1.5 to 2 minutes. In contrast, Mixtral-8*7B, despite utilizing an NVIDIA A100 GPU with 80GB of memory, took an average of 12 minutes to produce a single output for both Task 1 and each stage of Task 2. To maintain the integrity of the test conditions, any single output from Mixtral-8*7B that exceeded three hours were excluded from the analysis.

\subsection{Task 1: Classification Result}
In the evaluation of Psycho Analyst's performance, the 7-point likelihood scores provided by Psycho Analyst were binarized to classify subjects as either having or not having mental health issues. To determine the optimal threshold that would reflect the most accurate classification by the model, we tried threshold values from 3 to 7 (cut off at \textgreater{}= the threshold value) and model accuracy for the four backgound knowledge configurations across different score threshold values are shown in Fig. 3. The results indicate that score threshold 5 consistently yielded the highest model accuracy. Therefore, we used score 5 as the cut-off point to binarized the 7-point likelihood scores provided by Psycho Analyst in subsequent evaluations of Psycho Analyst's performance.
\begin{figure}[h]
  \centering
  \includegraphics[width=\columnwidth]{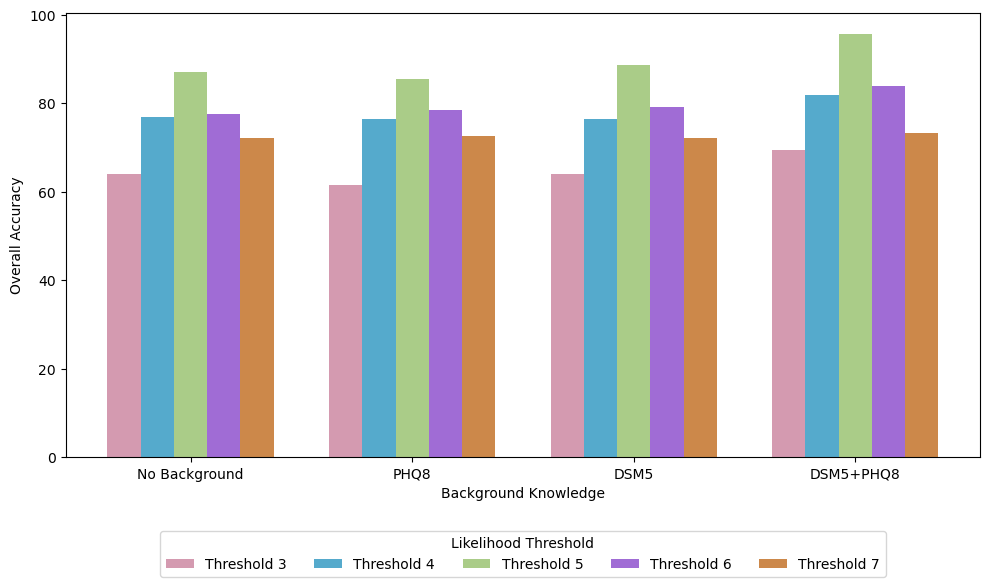} 
  \caption{Psycho Analyst prediction accuracy on the overall DAIC-WOZ dataset for the four different background knowledge configurations and likelihood score threshold values from 3 to 7.}
\end{figure}

\input{table5}
\input{table6}

Table~\ref{tab:table5} compares Psycho Analyst's performance across four configurations of background knowledge in the complete DAIC-WOZ dataset, assessing metrics such as F1 score, Macro-F1 score, Accuracy, Recall, Precision, and ROC-AUC score.

With no background knowledge, the model achieves an F1 score of 0.769 and an Accuracy of 0.872. Adding PHQ-8 slightly lowers all performance metrics, while incorporating DSM-5 alone enhances them, yielding an F1 score of 0.792 and an Accuracy of 0.888. The combination of DSM-5 and PHQ-8 results in the highest improvements, achieving an F1 score of 0.929, Accuracy of 0.957, Recall of 0.945, Precision of 0.912, and ROC-AUC of 0.968, highlighting the synergistic effect of these tools in boosting diagnostic accuracy and reliability.

Further insights are provided in Table~\ref{tab:table6}, which compares the performance of the Psycho Analyst model on the DAIC-WOZ Test Set across various configurations. This includes an enhanced configuration that integrates DSM-5, PHQ-8, a detailed data description from the dataset’s curator, and an expanded training set. Performance is also compared with two baseline models.

The GPT-4 model without background knowledge shows moderate performance, with an F1 score of 0.759 and an Accuracy of 0.851. Adding PHQ-8 decreases performance to an F1 score of 0.690 and an Accuracy of 0.808, while including DSM-5 enhances outcomes to an F1 score of 0.786 and an Accuracy of 0.872. The combined use of DSM-5 and PHQ-8 significantly boosts performance, similar to previous results, achieving an F1 score of 0.857 and Accuracy of 0.915. Additional enhancements from incorporating detailed data descriptions and an expanded training set further improve metrics, elevating the F1 score to 0.929 and Accuracy to 0.957.

Comparative analysis with GPT-4o and Mixtral-8*7B, both lacking background knowledge, highlights Psycho Analyst's superior performance. GPT-4o reaches an F1 score of 0.667 and accuracy of 0.702, while Mixtral-8*7B shows slightly lower scores, emphasizing Psycho Analyst’s advantages even without enhanced configurations.

These results indicate that the incorporation of diverse background knowledge, along with relevant data descriptions and training data, significantly enhances Psycho Analyst GPT's diagnostic capabilities. The improved performance demonstrates its potential as a robust tool for mental health pre-screening.

\subsection{Task 2: PHQ-8 Score Computation Result}

\begin{figure*}[h!]
  \centering
  \includegraphics[width=0.8\linewidth]{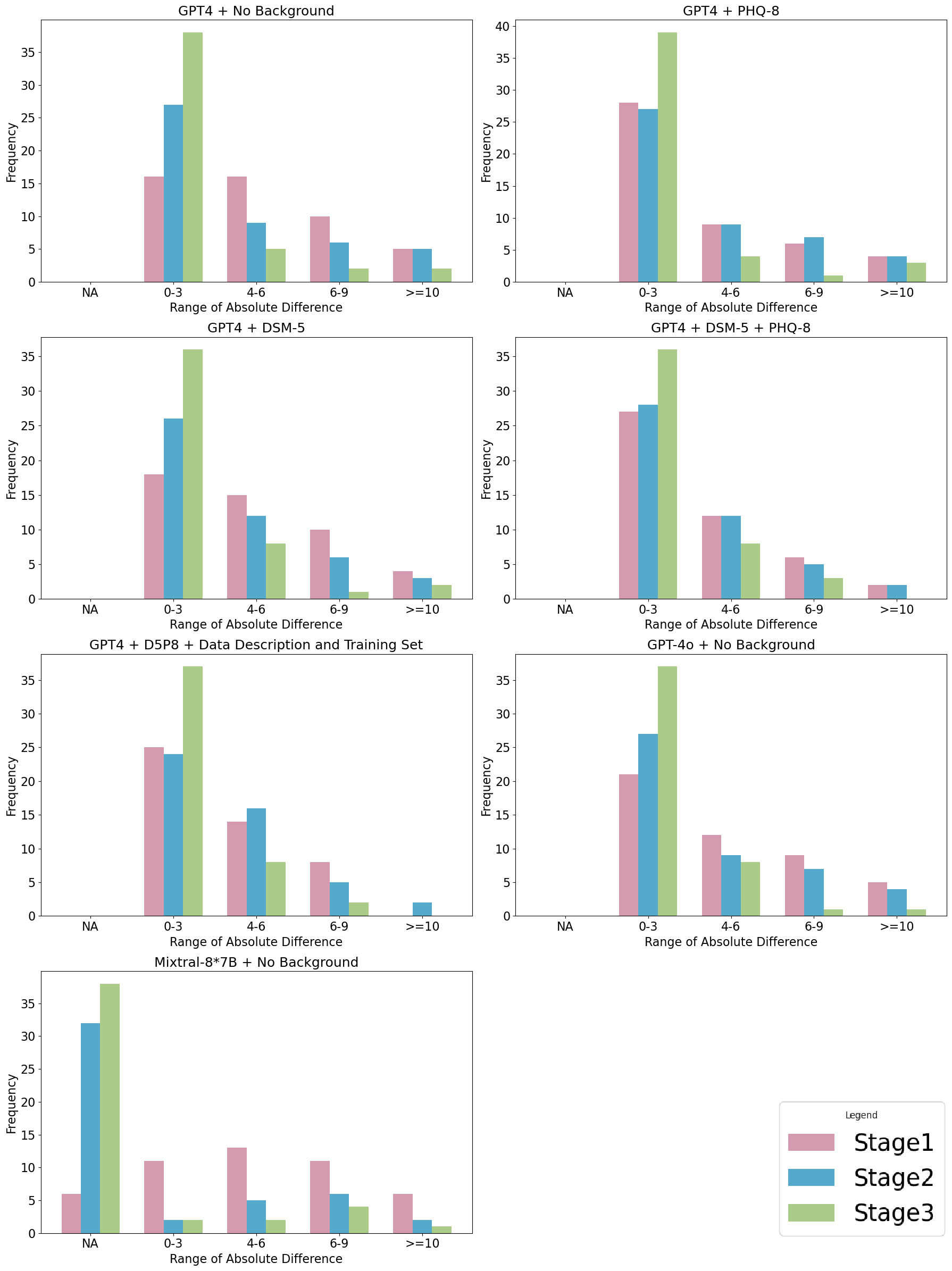}
  \caption{Frequency Distribution of Absolute Differences Across Three Stages of Evaluation for Various Models on the DAIC-WOZ Test Set.}
  \label{fig:output}
\end{figure*}

\input{MAEwide}

Figure \ref{fig:output} illustrates the distribution of absolute differences for the three-stage PHQ-8 score computation. The histograms depict the frequency of absolute differences between the true PHQ-8 scores and those estimated by various configurations of the GPT-4 model, the GPT-4o model, and the Mixtral-8*7B model. Each bar represents the frequency of a specific range of differences, demonstrating the accuracy of each configuration in accurately estimating mental health statuses.

When GPT-4 is enhanced with the PHQ-8 tool, both the initial assessment in Stage 1 and the detailed breakdown in Stage 2 exhibit similar performance, with minimal variation in the distribution of absolute differences. This consistency suggests that the inclusion of PHQ-8 provides a stable and reliable basis for both the initial analysis and the subsequent detailed scoring breakdown.

For other configurations of GPT models excluding PHQ-8, a noticeable improvement in performance during the detailed breakdown (Stage 2) is observed compared to the initial analysis (Stage 1). This indicates that a reassessment and detailed exploration of the initial model's assessment can significantly refine the accuracy of the models' evaluations, correcting any deficiencies noted in the initial analysis.

Moreover, in Stage 3, where other evaluators' opinions were provided as a reference, the GPT models still demonstrated a significant degree of autonomy. Despite the influence of external assessments, these models maintained a substantial level of self-decision, indicating their robustness in integrating and yet independently assessing complex clinical information.

The performance of Mixtral-8*7B is markedly different from the other models. A significant number of assessments were marked as 'NA' (not available), indicating that the model failed to complete the 3-stage task in many instances. Only a few cases reached all the way through Stage 3, suggesting substantial issues with data handling or model capability. This extensive occurrence of missing values undermines the value of Mixtral-87B's computations in this task, rendering it less valuable.

Table~\ref{tab:MAE} presents detailed metrics of these assessments, including Mean Absolute Error (MAE), Root Mean Square Error (RMSE), and R-squared values for each model configuration. The results indicate that the Psycho Analyst model equipped with both DSM-5 and PHQ-8 significantly outperforms other configurations. This performance is further enhanced when additional data descriptions and training data are included. This table allows for a comprehensive comparison of performance across all stages of the PHQ-8 score computation task.

\section{conclusion and discussion}
The results from our evaluation of the Psycho Analyst custom GPT model represent a significant advancement in the use of GPT technology for mental health pre-screening. The model’s enhanced performance, particularly when integrated with DSM-5, PHQ-8, detailed data descriptions, and robust training data, demonstrates its ability to handle nuanced language processing and its potential for precise mental health diagnostics, both in binary classification and PHQ-8 computation. The findings underscore that leveraging comprehensive background knowledge markedly improves the model's diagnostic capabilities, as evidenced by the notable increase in accuracy detailed in the tables and figures presented.

Our results also reveal that providing the PHQ-8 criteria as background enhances the stability of the initial PHQ-8 score computations. Conversely, for models without this background, prompting the model to explain its assessments significantly improves performance. Furthermore, the Psycho Analyst model demonstrates an ability to think independently, as evidenced by its occasional disagreements with other assessments. This independence is crucial for reliable and unbiased mental health evaluations.

The superior performance of the Psycho Analyst model, indicated by high F1, Macro-F1 scores, and low error metrics, confirms its reliability for initial mental health screenings. This reliability is vital in clinical settings where early detection can profoundly impact patient outcomes. By providing a robust tool for early diagnosis, Psycho Analyst could potentially streamline the diagnostic process, reduce the workload on mental health professionals, and enhance accessibility for patients. This model not only enhances diagnostic precision but also ensures that interventions are timely and contextually appropriate, significantly advancing the field of mental health care.

\input{final_table}

When compared to recent approaches in DAIC-WOZ transcript analysis, the Psycho Analyst model demonstrated superior performance not only in accuracy but also in functionality, as shown in Table~\ref{tab:final_table}.

Looking forward, the application of the Psycho Analyst model promises substantial benefits. Integrated into an AI-driven conversational chatbot familiar with mental health criteria, it enables at-risk populations to receive timely, low-cost, and barrier-free assessments at home. This capability is poised to significantly enhance public mental health outcomes by facilitating early detection and intervention. Furthermore, leveraging the advanced language capabilities of large language models (LLMs), the Psycho Analyst can also assist clinical professionals by providing a second opinion. This independent thinking ability can aid clinicians in self-checking their evaluations, contributing to more standardized and objective assessments in clinical settings.

Our next steps with the Psycho Analyst model involve expanding its capabilities to analyze social media data. This advancement will allow us to identify and understand mental health trends and indicators from the vast, unstructured data available online, in real time. Adopting this proactive approach to mental health surveillance promises to transform how we monitor well-being on a large scale. Additionally, we are in the process of developing a conversational interview chatbot. In conjunction with the Psycho Analyst, this tool will make mental health assessments more dynamic and personalized, enhancing accessibility and reducing the social pressure often associated with traditional screening methods.

\textbf{Future Data Privacy and Ethical Considerations:} Currently, our study leverages published, anonymous data, mitigating immediate concerns regarding patient confidentiality. However, as we look to the future application of Psycho Analyst in more diverse contexts, data privacy and model safety emerge as critical considerations. As the use of generative AI in mental health expands, ensuring the protection of sensitive information becomes paramount.

\textbf{Limitations:} Despite the promising results, the output from Psycho Analyst exhibited inconsistencies, particularly when processing large volumes of text, which led to occasional hallucinations. This highlights the need for further refinement of the model to ensure reliability and coherence in its responses. Addressing these limitations will be a primary focus of our continued research.

In conclusion, our research provides a vision of the future in mental health diagnostics and care, where generative AI tools like Psycho Analyst GPT play a crucial role. By narrowing the gap between technological innovation and healthcare requirements, such models have the potential to revolutionize mental health services, making them more accessible, efficient, and patient-focused.

\section*{Acknowledgment}
ChatGPT was used to revise the writing to improve the spelling, grammar, and overall readability.

\bibliographystyle{IEEEtran}
\bibliography{reference}

\end{document}

%% file: table1.tex
\begin{table}[t]
\caption{Number of participants in subsets of the DAIC-WOZ dataset and the numbers of participants with PHQ-8 scores \textless or \textgreater= 10.}
\label{table:table1}
\centering
\begin{tabular}{l|ccc}
\hline
\textbf{} & \textbf{Overall} & \textbf{Training } & \textbf{Test} \\
\hline
NoP (No. of Participants) & 187 & 105 & 47 \\

NoP with PHQ-8 Scores \textgreater=10 (A)& 55 & 30 & 14 \\

NoP with PHQ-8 Scores \textless10 (B)& 132 & 75 & 33 \\

Ratio A/B & 0.42 & 0.40 & 0.42 \\
\hline
\end{tabular}
\end{table}

%% file: table2.tex
\begin{table}[t]
\centering
\caption{Sample excerpts of raw transcript from the interactions in DAIC-WOZ dataset.}
\label{tab:table2}
\small
\begin{tabularx}{\columnwidth}{@{}lX@{}}
\toprule
Speaker     & Dialogue \\ \midrule
Ellie       & hi i'm ellie thanks for coming in today \\
Ellie       & i was created to talk to people in a safe and secure environment \\
Ellie       & think of me as a friend i don't judge i can't i'm a computer \\
Ellie       & i'm here to learn about people and would love to learn about you \\
Ellie       & i'll ask a few questions to get us started and please feel free to tell me anything your answers are totally confidential \\
Ellie       & how are you doing today \\
Participant & good \\
Ellie       & that's good \\
Ellie       & where are you from originally \\
Participant & atlanta georgia \\
Ellie       & really \\
Ellie       & why'd you move to l\_a \\
Participant & um my parents are from here um \\
Ellie       & how do you like l\_a \\
Participant & i love it \\
Ellie       & what are some things you really like about l\_a \\
Participant & i like the weather \\
Participant & i like the opportunities \\
Participant & um \\
Participant & yes \\ \bottomrule
\end{tabularx}
\end{table}

%% file: table3.tex
\begin{table}[t]
\centering
\caption{Sample Raw Participant Dialogue Compilation Segment}
\label{tab:table3}
\begin{tabularx}{\columnwidth}{@{}lX@{}}
\toprule
Participant ID & Joint Participant Dialogue Text \\
\midrule
300 & good./ atlanta georgia./ um my parents are from here... \\
301 & thank you./ mmm k./ i'm doing good thank you./ i'm ... \\
302 & i'm fine how about yourself./ i'm from los angeles... \\
303 & okay how 'bout yourself./ here in california. yeah. ... \\
304 & i'm doing good um./ from los angeles california./ ... \\
305 & i'm doing alright./ uh originally i'm from california./ uh... \\
306 & fine./ uh colorado./ mhm./ uh career./ career ... \\
307 & \textless laughter\textgreater./ um moscow./ um my family moved to... \\
308 & los angeles california./ yes./ um the southern... \\
309 & \textless laughter\textgreater./ \textless laughter\textgreater yeah\textless laughter\textgreater./ ... \\
\bottomrule
\end{tabularx}
\end{table}

%% file: table4.tex
\begin{table}[t!]
\centering
\caption{Sample Psycho Analyst GPT output.}
\label{tab:table4}
\begin{tabularx}{\columnwidth}{@{}X@{}}
\toprule
To assess the mental health status of each participant, I will analyze their responses using my knowledge of psychological patterns and criteria associated with mental health issues, specifically focusing on signs of anxiety, depression, stress, and bipolar disorder. It's important to remember that these assessments are not definitive diagnoses but rather interpretations based on the provided text. Let's proceed with each participant.

\textbf{Participant 300:}
Mental Health Assessment: The participant seems to exhibit a balanced outlook on life, with no overt signs of distress in their responses. They mention feeling frustrated occasionally but also engage in positive activities and have supportive family relationships.
Likelihood of Ongoing Mental Health Disorder: \textbf{2/7}. The participant appears to have some minor irritations but overall demonstrates a healthy psychological profile.

\textbf{Participant 306:}
Mental Health Assessment: The participant discusses past challenges, including a legal issue and the implications of peer pressure on their child. They also mention feeling stressed and fatigued but do not exhibit strong signs of depression or severe anxiety.
Likelihood of Ongoing Mental Health Disorder: \textbf{3/7}. While there are signs of stress and past trauma, there's not enough evidence to suggest a severe ongoing mental health disorder.

\textbf{Participant 308:}
Mental Health Assessment: This participant displays several signs of depression, including joblessness, feelings of hopelessness, sleep disturbances, and a history of therapy for mood issues. The recent loss of a family member and homelessness add to the severity of their situation.
Likelihood of Ongoing Mental Health Disorder: \textbf{6/7}. The participant's current life circumstances and psychological state strongly indicate ongoing depression.

\textbf{Participant 309:}
Mental Health Assessment: The participant shows variability in mood and mentions a history of depression. They also speak about anger management issues and recent life changes (wife leaving). This fluctuation in mood and emotional response could indicate bipolar tendencies or severe stress.
Likelihood of Ongoing Mental Health Disorder: \textbf{5/7}. The participant's history and current emotional state suggest a significant possibility of ongoing mental health issues, possibly bipolar disorder or depression.

\textbf{Participant 311:}
Mental Health Assessment: The participant talks about experiences with depression, hospitalization for mental health issues, and auditory hallucinations. They also mention ongoing struggles with sleep and mood. These are strong indicators of a serious mental health condition, likely chronic in nature.
Likelihood of Ongoing Mental Health Disorder: \textbf{7/7}. The participant's history and current symptoms strongly suggest an ongoing and severe mental health disorder. 

\textit{These assessments are based on the information provided and should be taken as preliminary insights. It's crucial for individuals showing signs of mental health issues to seek professional help for accurate diagnosis and appropriate treatment.}
\\
\bottomrule
\end{tabularx}
\end{table}

%% file: table_3stage.tex
\begin{table*}[ht]
\caption{Sample 3-Stage Assessment Output}
\label{tab:table_3stage}
\centering
\begin{tabular}{c|p{0.8\textwidth}}
\toprule
\textbf{Stage} & \textbf{Output} \\
\hline
Stage 1 & \textbf{Signs of Symptoms:}\\
 & \textit{Motivation and Interest:} The individual frequently expresses a lack of motivation and interest in previously enjoyed activities, stating: ``I find it hard to relax, I don’t have the same outlets that I did before... I’ve lost my motivation.''\\
 & \textit{Feelings of Hopelessness:} There are multiple indications of feeling down and hopeless: ``I've kind of lost myself... I don’t really have anybody in my life who I think cares enough about me.''\\
 & \textit{Fatigue and Energy:} Persistent fatigue and low energy are noted: ``I get inspired every once in a while but overall I just don’t seem to be making much progress.''\\
 & \textit{Self-Worth:} Thoughts of worthlessness or feeling like a failure are prevalent: ``I don’t really have any pursuits... I don’t have much of a sense of pride anymore.''\\
 & \textit{Concentration:} Difficulty concentrating is evident: ``I’ve lost quite a bit of that... whatever. I used to... help people and now I don't seem to be able to help myself.''\\
 & \textbf{Estimated PHQ-8 Score:} 2+3+1+2+1+2+1+1=13 (indicating moderate depressive symptoms)\\
 & \textbf{Summary:} The individual exhibits substantial symptoms of depression, with significant impairment in social and occupational areas. Based on the transcript and using the DSM-5 and PHQ-8 criteria, the likelihood of an ongoing mental health disorder is high. Professional mental health support is recommended for a comprehensive evaluation and appropriate intervention. \\
\hline
Stage 2 & \textbf{PHQ-8 Scoring Breakdown:}\\
 & \textit{Little interest or pleasure in doing things:} ``I used to do a lot of things lately... I find it hard to relax. I don't have the same outlets that I did before. I don't pursue them as much anymore.'' \textbf{Score:} 2 (More than half the days)\\
 & \textit{Feeling down, depressed, or hopeless:} ``I'm depressed about it... There it is right there.'' and ``I've kind of lost myself... I used to think I had a purpose and such, but I don't really have anybody in my life who I think cares enough about me.'' \textbf{Score:} 3 (Nearly every day)\\
 & \textit{Trouble falling or staying asleep, or sleeping too much:} ``Yeah, when I get a chance to fall asleep, I don't have any problem with it.'' \textbf{Score:} 0 (Not at all)\\
 & \textit{Feeling tired or having little energy:} ``I've lost quite a bit of that motivation.'' and ``I get inspired every once in a while but overall I just don't seem to be making much progress.'' \textbf{Score:} 2 (More than half the days)\\
 & \textit{Feeling bad about yourself, or that you are a failure, or have let yourself or your family down:} ``I don't really have any pursuits... I don't have much of a sense of pride anymore.'' and ``I've lost my motivation... I'm depressed about it, you know.'' \textbf{Score:} 2 (More than half the days)\\
 & \textit{Trouble concentrating on things, such as reading the newspaper or watching television:} ``I don't seem to be able to help myself.'' \textbf{Score:} 1 (Several days)\\
 & \textbf{Total PHQ-8 Score:} 2+3+0+2+0+2+1+0 = 10\\
 & \textbf{Summary:} The individual exhibits symptoms consistent with moderate depression, as indicated by a PHQ-8 score of 10. This suggests a moderate to high probability of an ongoing depressive disorder. Professional evaluation and intervention are recommended.\\
\hline
Stage 3 & \textbf{Re-analysis of PHQ-8 Scoring :}\\
 & Let's reassess the PHQ-8 score in light of the other evaluator's assigned value of 5 and compare it to my initial assessment....\\
 & \textbf{Discussion:} Comparing this to the score of 5 assigned by another evaluator: The transcript reflects significant depressive symptoms, particularly in areas of interest, mood, energy, and self-worth, which seem to align more closely with a higher score than 5. The initial analysis of symptoms and their frequency suggests a higher PHQ-8 score.\\
 & \textbf{Conclusion:} The revised PHQ-8 score of 10 is consistent with the symptoms described in the transcript. This suggests moderate depressive symptoms. Therefore, the likelihood rating of 5 (moderate probability of an ongoing depressive disorder) is appropriate based on this re-evaluation. In conclusion, I \textbf{do not agree} with the assessment of a PHQ-8 score of 5 by another evaluator. The re-evaluated score of 10 aligns better with the individual's reported symptoms. This highlights the importance of thorough and consistent evaluation of the individual's experiences against established criteria.\\

\bottomrule
\end{tabular}
\end{table*}

%% file: table5.tex
\begin{table*}[t]
\centering
\small
\caption{Performance comparison of Psycho Analyst on the overall DAIC-WOZ dataset with various background knowledge configurations}
\label{tab:table5}
\begin{tabular}{l|ccccccc}
\toprule
Background Knowledge & F1 & Macro-F1 & Accuracy & Recall & Precision & ROC-AUC \\
\midrule
GPT4 + No Background & 0.769 & 0.840 & 0.872 & 0.727 & 0.816 & 0.853 \\

GPT4 + PHQ-8 & 0.738 & 0.819 & 0.856 & 0.691 & 0.792 & 0.851 \\

GPT4 + DSM-5 & 0.792 & 0.858 & 0.888 & 0.727 & 0.870 & 0.882 \\

GPT4 + DSM-5 + PHQ-8 & \textbf{0.929} & \textbf{0.949} & \textbf{0.957} & \textbf{0.945} & \textbf{0.912} & \textbf{0.968} \\

\bottomrule
\end{tabular}
\end{table*}

%% file: table6.tex
\begin{table*}[t]
\centering
\small
\caption{Performance comparison of Psycho Analyst on the DAIC-WOZ Test Set with Various Configurations and Models}
\label{tab:table6}
\begin{threeparttable}
\begin{tabular}{l|cccccc}
\toprule
Models and Background Knowledge &    F1 &  Macro-F1 &  Accuracy &  Recall &  Precision &  ROC-AUC \\
\midrule
GPT4 + No Background & 0.759 & 0.825 & 0.851 & 0.786 & 0.733 & 0.827 \\
GPT4 + PHQ-8 & 0.690 & 0.776 & 0.808 & 0.714 & 0.667 & 0.804 \\ 
GPT4 + DSM-5 & 0.786 & 0.847 & 0.872 & 0.786 & 0.786 & 0.860 \\ 
GPT4 + DSM-5 + PHQ-8 & \textbf{0.857} & \textbf{0.898} & \textbf{0.915} & \textbf{0.857} & \textbf{0.857} & \textbf{0.905} \\
GPT4 + DSM-5 + PHQ-8 + Data Description and Training Set & \textbf{0.929} & \textbf{0.949} & \textbf{0.957} & \textbf{0.929} & \textbf{0.929} & \textbf{0.952} \\ \hline
GPT-4o + No Background & 0.667 & 0.699 & 0.702 & 0.1.00 & 0.500 & 0.838 \\
Mixtral-8*7B  + No Background & 0.649 & 0.680 & 0.683 & 0.923 & 0.500 & 0.780 \\
\bottomrule
\end{tabular}
\begin{tablenotes}
\small
\item Note: The results for Mixtral-8*7B were computed after excluding 6 missing values.
\end{tablenotes}
\end{threeparttable}
\end{table*}

%% file: MAEwide.tex
\begin{table*}[htbp]
\centering
\caption{Comparative Performance of 'Psycho Analyst' in Three-Stage PHQ-8 Analyses with Various Configurations Against Baseline Models GPT-4o and Mixtral-8*7B on the DAIC-WOZ Test Set}
\label{tab:MAE}
\begin{threeparttable}
\begin{tabular}{lccccccccc}
\toprule
 & \multicolumn{3}{c}{Stage1} & \multicolumn{3}{c}{Stage2} & \multicolumn{3}{c}{Stage3} \\
\cmidrule(lr){2-4} \cmidrule(lr){5-7} \cmidrule(lr){8-10}
Model and BackgroundKnowledge & MAE & RMSE & R-squared & MAE & RMSE & R-squared & MAE & RMSE & R-squared \\
\midrule
GPT4 + No Background & 4.42 & 5.59 & 0.24 & 3.53 & 5.00 & 0.39 & 1.57 & 2.72 & 0.82 \\
GPT4 + PHQ-8 & 3.28 & 4.39 & 0.53 & 3.32 & 4.49 & 0.51 & 1.68 & 2.95 & 0.79 \\
GPT4 + DSM-5 & 3.98 & 5.22 & 0.33 & 3.00 & 4.19 & 0.57 & 1.57 & 2.78 & 0.81 \\
GPT4 + DSM-5 + PHQ-8 & 3.04 & 4.16 & 0.58 & \textbf{2.91} & 3.98 & 0.61 & 1.64 & 2.59 & 0.84 \\
GPT4 + D5P8 + Data Description and Training Set  & \textbf{2.89} & \textbf{3.69} & \textbf{0.67} & 2.96 & \textbf{3.81} & \textbf{0.64} & 1.53 & 2.26 & 0.88 \\
GPT-4o + No Background & 4.04 & 5.40 & 0.29 & 3.45 & 4.78 & 0.44 & 1.29 & 2.44 & 0.85 \\
Mixtral 8*7B & 4.88 & 6.11 & 0.14 & 5.93 & 7.35 & -0.50 & 5.44 & 6.94 & 0.00  \\
\bottomrule
\end{tabular}
\begin{tablenotes}
\small
\item * D5P8 is DSM-5 + PHQ-8.
\item * A negative R-squared value indicates that the model performs worse than a model that simply predicts the mean of the dependent variable for all observations, suggesting poor model fit or specification.
\item * For Mixtral 8*7B, all metrics were calculated after removing missing values.
\end{tablenotes}
\end{threeparttable}
\end{table*}

%% file: final_table.tex
\begin{table*}[t]
\caption{Functionality comparison of Psycho Analyst Custom GPT with previous methods on DAIC-WOZ Transcript Analysis}
\label{tab:final_table}
\centering
\small
\begin{tabular}{l|cccc}
\hline
\textbf{Approach Comparison} & \textbf{High Accuracy} & \textbf{Reliability} & \textbf{Lexicon Analyse} & \textbf{Provide Explanation} \\ \hline
This Work  & \checkmark & \checkmark & \checkmark & \checkmark \\
Yadav et al. (2023) \cite{yadav2023novel}  & \checkmark & \checkmark &   \\
Belser et al. (2023) \cite{belser2023comparison}  &  & \checkmark &    \\
Villatoro et al. (2021) \cite{villatoro2021approximating} &  & \checkmark & \checkmark &    \\
\bottomrule
\end{tabular}
\end{table*}